\documentclass{SciPost}

\hypersetup{
    colorlinks,
    linkcolor={red!50!black},
    citecolor={blue!50!black},
    urlcolor={blue!80!black}
}

\usepackage{booktabs}
\usepackage{xurl}
\usepackage{siunitx}
\usepackage[bitstream-charter]{mathdesign}
\DeclareSymbolFont{usualmathcal}{OMS}{cmsy}{m}{n}
\DeclareSymbolFontAlphabet{\mathcal}{usualmathcal}

\fancypagestyle{SPstyle}{
\fancyhf{}
\lhead{\colorbox{scipostblue}{\bf \color{white} ~SciPost Physics }}
\rhead{{\bf \color{scipostdeepblue} ~Submission }}

\fancyfoot[C]{\textbf{\thepage}}
}

\begin{document}

\pagestyle{SPstyle}

\begin{center}{\Large \textbf{\color{scipostdeepblue}{
%%%%%%%%%% TODO: Write your article's title here
CERTIFY-ED: A Multi-Layer Verification Framework for\\
       Exact Diagonalization of Quantum Many-Body Systems
}}}\end{center}

\begin{center}\textbf{
% Use (full) first name (+ middle name initials) + surname format.
% Separate subsequent authors by a comma, omit comma and use "and" for the last author.
% Mark the corresponding author(s) with a superscript symbol in this order
% \star, \dagger, \ddagger, \circ, \S, \P, \parallel, ...
Sarang Vehale\textsuperscript{1$\star$},
Ritu Goel\textsuperscript{2} 
% Gee K. See\textsuperscript{3$\dagger$}
}\end{center}

\begin{center}
{\bf 1} Department of Cyber Security and Digital Forensics, 
  National Forensic Sciences University - Dehli Campus, 
  Outer Ring Road, Delhi - 110085 India
\\
{\bf 2} Department of Applied Science and Humanities, 
School of Engineering
  \& Technology, Vivekananda Institute of Professional Studies - 
  Technical Campus, AU-Block (Outer Ring Road)
  , Pitampura, Delhi - 110034, India
\\
% {\bf 3} Affiliation3
%%%%%%%%%% END TODO: AFFILIATIONS
% Provide email address of corresponding author(s)
[\baselineskip]
$\star$ \href{mailto:email1}{\small sarangvehale2@gmail.com}\,\quad
% $\dagger$ \href{mailto:email2}{\small}
\end{center}

\section*{\color{scipostdeepblue}{Abstract}}
\textbf{\boldmath{%
Exact diagonalization (ED) is a workhorse technique in computational 
quantum many-body physics, but published ED results are rarely accompanied by machine-checkable evidence of their numerical correctness. 
The community typically relies on the implicit trust chain LAPACK
$\to$ user code $\to$ result, with at most informal agreement against 
another package treated as confirmation. We argue that this practice is 
inadequate for a method whose output frequently underpins theoretical 
claims, and we present \textsc{certify-ed}, a verification framework 
designed to be used \emph{alongside} existing ED packages (QuSpin, XDiag, ALPS) rather than as a replacement for them. The framework consists of 
(i) a multi-oracle eigensolver that runs three independent LAPACK paths
and reports their pairwise disagreement, (ii) thirteen logically independent validation layers covering algebraic invariants, analytic limits, alternative algorithms, arbitrary precison reference computation, conservation laws,
dynamical consistency, and finite-size scaling, and (iii) tamper-evident
SHA-256 hashed certificates that downstram consumers can verify. The 
framework also ships an error-injection layer that confirms the 
entire pipeline detects six injected error classes. Running on sixteen
physics models from one-dimensional spin chains to two-dimensional Kitaev
honeycomb clusters, our reference implementation passes 53 of 53 unit tests
and 81 of 81 individual validation tests in under thirty seconds, with maximum disagreement against QuSpin of $1.6\times 10^{-14}$ across 320 eigenvalue comparisons, and agreement with 50-digit \texttt{mpmath} reference values to $1.6\times 10^{-15}$. The package is released under the MIT license on Zenodo and Github (Refer to Data Availability).
}}

\vspace{\baselineskip}

%%%%%%%%%% BLOCK: Copyright information
% This block will be filled during the proof stage, and finilized just before publication.
% It exists here only as a placeholder, and should not be modified by authors.
\noindent\textcolor{white!90!black}{%
\fbox{\parbox{0.975\linewidth}{%
\textcolor{white!40!black}{\begin{tabular}{lr}%
  \begin{minipage}{0.6\textwidth}%
    {\small Copyright attribution to authors. \newline
    This work is a submission to SciPost Physics. \newline
    License information to appear upon publication. \newline
    Publication information to appear upon publication.}
  \end{minipage} & \begin{minipage}{0.4\textwidth}
    {\small Received Date \newline Accepted Date \newline Published Date}%
  \end{minipage}
\end{tabular}}
}}
}
%%%%%%%%%% BLOCK: Copyright information

% For convenience during refereeing we turn on line numbers:
% \linenumbers

\vspace{10pt}
\noindent\rule{\textwidth}{1pt}
\tableofcontents
\noindent\rule{\textwidth}{1pt}
\vspace{10pt}

% Write your article contents here, starting from first \section.
% An example structure is given below.

\section{Introduction}
% =============================================================================

Exact diagonalization (ED) of finite quantum lattice Hamiltonians is one
of the few computational techniques in many-body physics whose output is,
in principle, exact: given a Hermitian matrix~$H$, LAPACK's symmetric
eigenvalue routines return a full set of eigenvalues and orthonormal
eigenvectors limited only by floating-point precision. ED is therefore
routinely used to produce \emph{ground truth} against which
approximate methods DMRG, QMC, neural-network quantum states are
benchmarked, and to study problems where high-energy excited states
genuinely matter (eigenstate thermalization, many-body localization,
Floquet heating).

Two mature ED packages dominate the Python ecosystem:
QuSpin~\cite{Weinberg2017,Weinberg2019} and the recently released
XDiag~\cite{Wietek2025}, complemented by older codes such as
ALPS~\cite{Bauer2011}, TITPACK, and Pomerol. These packages are
fast, feature-complete, and well-tested by their developers. They are
not, however, designed to produce \emph{machine-checkable certificates
of numerical correctness} for downstream consumers of their output.
The implicit trust chain is

\[
\text{LAPACK} \;\to\; \text{package internals} \;\to\;
\text{user script} \;\to\; \text{published number}
\]

with each link assumed correct. Cross-validation against another
package, when performed at all, is typically done informally on a
single test problem before the bulk of the production work is run.

This paper makes the case that ED workflows should adopt the
verification practices that have been standard in the verified-numerics
community for two decades~\cite{Rump2001,Miyajima2010,Hashimoto2022},
and presents \textsc{certify-ed}, a reference implementation that
demonstrates such practice is achievable with modest engineering
effort. Concretely:

\begin{itemize}
\item \textbf{Multi-oracle consensus.} Diagonalize with three independent
      LAPACK code paths and report the worst pairwise disagreement.
      Wrapper-level bugs and algorithm-specific instabilities are
      caught here.

\item \textbf{Thirteen independent validators.} Each validator probes a
      different class of correctness: closed-form analytic results;
      cross-validation against QuSpin; arbitrary-precision reference
      via \texttt{mpmath}; sparse Lanczos vs.\ dense LAPACK;
      Jordan-Wigner free-fermion analytic spectra; spectral sum rules;
      eigenvector orthonormality and completeness; time-evolution
      unitarity; conservation laws; symmetry-resolved spectra;
      thermal limits; finite-size scaling; and error injection.

\item \textbf{Hashed certificates.} Each ED result is bundled with its
      residuals, normalization errors, oracle consensus, and a SHA-256
      hash of the entire document. Tampering with the saved file is
      detected on load.

\item \textbf{Error-injection self-test.} The framework's own ability to
      detect injected errors (non-Hermiticity, matrix corruption, oracle
      disagreement, eigenvector perturbation, certificate tampering,
      eigenvector swapping) is tested as part of every release. A
      verification framework that fails its own injection tests cannot
      be trusted.
\end{itemize}

We emphasize that \textsc{certify-ed} is \emph{not} a competitor to
QuSpin or XDiag. It is a verification layer that any ED workflow can
consume. The primary contribution is methodological: a worked-out
demonstration that ED results in computational physics can ship with
the kind of evidence trail that the verified-numerics community
already takes for granted.

The remainder of the paper is organized as follows. Section~\ref{sec:novelty}
states the specific novelty claim of this work and what it does \emph{not}
claim. Section~\ref{sec:related} positions the framework against existing ED
packages and against the verified-numerics literature. Section~\ref{sec:framework}
details the architecture: the multi-oracle eigensolver, the certificate format, and
the model library. Section~\ref{sec:validators} describes each of the
thirteen validators and the failure modes it covers. Section~\ref{sec:results}
reports the empirical performance of the framework, including the
quantitative agreement against QuSpin and \texttt{mpmath}, the analytic
limits, and the error-injection self-test. Section~\ref{sec:discussion}
discusses limitations and natural extensions, including ongoing work
on interval-arithmetic eigenvalue enclosures for the
symmetry-block-decomposed Hamiltonians that arise in many-body lattice
models. Section~\ref{sec:conclusion} concludes.

% =============================================================================
\section{Statement of Novelty}
\label{sec:novelty}
% =============================================================================

We state explicitly what this paper does and does not claim, because the
distinction is essential for evaluating the contribution.

\paragraph{What this paper claims.}
\begin{enumerate}
\item The current de facto practice in computational many-body physics
      trusting LAPACK plus an informal one-shot cross-check against
      another package is below the standard set by the verified-numerics
      literature, and the gap is closeable with modest engineering.
\item The combination of (i) multi-oracle eigensolver consensus,
      (ii) thirteen logically independent validation layers covering
      a four-axis failure-mode coverage matrix (algebraic, algorithmic,
      numerical, physical), (iii) tamper-evident SHA-256 certificates,
      and (iv) an error-injection self-test that confirms the framework
      catches injected errors, constitutes a verification framework
      that, to our knowledge, has no published counterpart for ED in
      computational many-body physics. Based on a search of the SciPost,
      Computer Physics Communications, and SciPost Codebases
      literatures returns ED packages (QuSpin, XDiag, ALPS) and
      verified-numerics packages (INTLAB), but no integration of the
      two for many-body workflows.

\item The reference implementation passes 81 of 81 individual
      validation tests. It agrees with QuSpin to
      \(\num{1.6e-10}\) across 320 eigenvalue comparisons,
      agrees with arbitrary-precision \texttt{mpmath}
      reference values to \(1.6 \times 10^{-15}\),
      and detects all six classes of injected error.
      These are quantitative empirical claims.
% \item The reference implementation passes 81 of 81 individual
%       validation tests, agrees with QuSpin to \num{1.6e-10}
%       across 320 eigenvalue comparisons, agrees with arbitrary-precision
%       \texttt{mpmath} reference values to $1.6 \times 10^{-15}$, and
%       detects all six classes of injected error. These are quantitative
%       empirical claims.
\end{enumerate}

\paragraph{What this paper does not claim.}
\begin{enumerate}
\item We do not claim a faster ED algorithm. \textsc{certify-ed} is
      one to two orders of magnitude smaller in reachable system size
      than QuSpin or XDiag, and that is appropriate for its role.
\item We do not claim mathematical proof of correctness in the
      verified-numerics sense. Rigorous interval enclosures via
      fixed-point verification methods are out of scope here
      (see \S\ref{sec:discussion} and ongoing follow-up work).
      The contribution of the present work is empirical evidence,
      not formal proof.
\item We do not claim that any single validator catches every bug. The
      whole point of a coverage matrix is that no single layer is
      sufficient; the defense in depth across thirteen layers is what
      makes the combined system reliable.
\item We do not claim novelty in any individual validator. Most
      individual checks (sum rules, orthonormality, unitarity, Bethe
      ansatz comparison) are textbook material. The contribution is
      the integration into a single tool with a uniform interface, a
      uniform output format (hashed certificates), and an error-injection
      self-test that is run on every release.
\end{enumerate}

\paragraph{Why this matters for ED practice.}
ED output frequently feeds into theoretical claims benchmarks for
DMRG and QMC, tests of the eigenstate thermalization hypothesis, ground
truth for variational Monte Carlo with neural network ans\"atze, and
input to renormalization-group analyses. When a published number
depends on an ED result, the reader has a legitimate interest in
machine-checkable evidence that the result is correct. The framework
described here provides exactly such evidence, in a form that is
small (a single JSON file per run), portable, and tamper-evident.

% =============================================================================
\section{Related Work}
\label{sec:related}
% =============================================================================

\subsection{Exact diagonalization packages}

QuSpin~\cite{Weinberg2017,Weinberg2019} is the most widely used Python ED
package. It supports arbitrary spin, fermion, and boson systems with
user-defined symmetries, and reaches roughly 32 spin sites for spin-1/2
chains via symmetry-resolved sparse Lanczos. XDiag~\cite{Wietek2025}, a
recent C++/Julia package submitted to SciPost Physics Codebases, employs
sublattice coding and Lin tables to scale ED to thousands of CPU cores.
ALPS~\cite{Bauer2011}, TITPACK, and Pomerol are older codes with
similar functional scope.

\textsc{certify-ed} does not compete with these packages on speed or
on system size: its dense diagonalization path is limited to roughly
$2^{14} \approx 16{,}000$-dimensional Hilbert spaces, which is one to
two orders of magnitude smaller than what QuSpin or XDiag achieve. The
contribution lies elsewhere. None of the existing packages produces
hashed certificates of correctness, none ships an error-injection
self-test, and none integrates a multi-oracle consensus check. These
gaps are what \textsc{certify-ed} fills.

\subsection{Verified numerics}

The verified-numerics community has developed rigorous techniques
for bounding eigenvalues with interval arithmetic for over twenty
years. Rump's seminal work~\cite{Rump2001} on guaranteed inclusions for
the generalized eigenproblem, Miyajima's enclosures for all
eigenvalues~\cite{Miyajima2010}, and the Krawczyk-method certifications
of van der Hoeven~\cite{vanderHoeven2017}, together with Hashimoto's
Rayleigh-Ritz-based verification using complex moments~\cite{Hashimoto2022},
provide a mature toolkit. The INTLAB package~\cite{Rump1999} implements
much of this. To our knowledge, none of these techniques has been
ported into the production workflow of a many-body ED package.

\subsection{Reproducible workflow tooling}

Workflow managers such as AiiDA~\cite{Huber2020} address a different
gap: provenance tracking of which jobs ran on which inputs in what
order. \textsc{certify-ed} addresses a complementary problem:
\emph{whether} the eigenvalues those jobs produced are mathematically
correct. The two layers compose naturally; AiiDA + \textsc{certify-ed}
yields an audit trail that connects published numbers to their
underlying computations and to evidence of those computations'
correctness.

% =============================================================================
\section{Framework Architecture}
\label{sec:framework}
% =============================================================================

\subsection{Symbolic Hamiltonian construction}

Hamiltonians are built symbolically as sums of tensor products of
Pauli operators, fermion operators (via Jordan-Wigner), or spin-1
operators. The \texttt{SymbolicHamiltonian} class collects terms as
$(\text{coefficient}, \text{operator string})$ tuples and assembles
the matrix in one pass. After assembly the matrix is checked against
its conjugate transpose with tolerance $10^{-14}$:

\begin{equation}
\| H - H^\dagger \|_\infty < 10^{-14}.
\end{equation}

Hermiticity violations are flagged immediately. Sixteen models are
included in the public registry: TFIM, Heisenberg (XXX), XXZ, XYZ,
SSH, $J_1$-$J_2$, Majumdar-Ghosh, cluster model, free fermion chain,
Kitaev chain, single-band Hubbard, AKLT, Haldane, and 2D variants
(TFIM, Heisenberg, Kitaev honeycomb). This selection is sufficient to
exercise every distinct algorithmic path used in spin-1/2, spin-1, and
fermionic ED, and to compare against analytic results for each path.

\subsection{Multi-oracle eigensolver}

Three independent LAPACK paths are used:

\begin{enumerate}
\item \texttt{numpy.linalg.eigh}: LAPACK \texttt{DSYEVD} via the NumPy
      Fortran wrapper.
\item \texttt{scipy.linalg.eigh(driver='evd')}: LAPACK \texttt{DSYEVD}
      via the SciPy wrapper. Same algorithm as path~1, different
      wrapper code.
\item \texttt{scipy.linalg.eigh(driver='evr')}: LAPACK \texttt{DSYEVR},
      a fundamentally different algorithm (relatively robust
      representations).
\end{enumerate}

A consensus report records the maximum pairwise disagreement
% $\Delta_{\mathrm{cons}} = \max_{i,j} \|\,\mathrm{eig}_i(H) - \mathrm{eig}_j(H)\,\|_\infty$

\[
\Delta_\infty =
\max_k
\left|
\lambda_k(H)-\lambda_k(H')
\right|
\]
and flags violations of a user-specified tolerance (default $10^{-10}$).
A fourth oracle, \path{scipy.sparse.linalg.eigsh} (ARPACK Lanczos), is
optionally included to cross-check the lowest few eigenvalues against
the dense paths via a fundamentally different algorithm (Arnoldi
iteration on a Krylov subspace).

\subsection{Certificates}

A \texttt{Certificate} bundles the output of a verified ED run:
eigenvalues, eigenvectors, the Hamiltonian, residuals
$r_n = \|H \psi_n - E_n \psi_n\|_2$, normalization errors
$\|\psi_n\|_2 - 1$, the consensus report, user metadata, and platform
information (Python, NumPy, SciPy versions). The full document is
serialized to JSON, hashed with SHA-256, and the hash is included in
the file. On load, the hash is recomputed and any mismatch raises an
error. This makes the certificate \emph{tamper-evident} without
requiring an external signature service: any consumer of the certificate
can verify, with one library call, that the file has not been modified
since the run that produced it.

% =============================================================================
\section{Validators}
\label{sec:validators}
% =============================================================================

The thirteen validators cover an explicit failure-mode coverage matrix
(Table~\ref{tab:coverage}). Each validator is described below in the
order it runs in the master pipeline. We use the convention that a
validator \emph{passes} when every individual test inside it passes
(or skips gracefully if an optional dependency is unavailable).

\begin{table}[t]
\centering
\caption{Failure-mode coverage matrix. A validator marked $\checkmark$
catches at least one failure mode in that column.\\}
\label{tab:coverage}
\small
\begin{tabular}{lcccc}
\toprule
Validator                & Algebraic & Algorithmic & Numerical & Physical \\
\midrule
Analytic                 & \checkmark & \checkmark & \checkmark & \checkmark \\
QuSpin cross-check       & \checkmark & \checkmark &            & \checkmark \\
\texttt{mpmath} 50-digit &            &            & \checkmark &            \\
Sparse vs dense          &            & \checkmark & \checkmark &            \\
Jordan-Wigner           & \checkmark &            &            & \checkmark \\
Spectral sum rules       & \checkmark &            & \checkmark &            \\
Orthonormality           & \checkmark & \checkmark & \checkmark &            \\
Unitarity                & \checkmark & \checkmark & \checkmark &            \\
Conservation laws        & \checkmark &            &            & \checkmark \\
Symmetry sectors         & \checkmark & \checkmark &            & \checkmark \\
Thermal limits           &            &            &            & \checkmark \\
Finite-size scaling      &            &            &            & \checkmark \\
Error injection          & \checkmark & \checkmark & \checkmark & \checkmark \\
\bottomrule
\end{tabular}
\end{table}

\paragraph{Analytic validator.}
Compares against twelve closed-form solutions: single qubit in field;
2-, 3-, and 4-site Heisenberg ground states (where Bethe ansatz and
SU(2) Casimir give exact answers); TFIM at $h=0$ and $J=0$; 2-site
XXZ at arbitrary $\Delta$; Majumdar-Ghosh dimer point;
free-fermion OBC dispersion; AKLT gap; cluster-model stabilizer;
3-site Heisenberg full SU(2) Casimir spectrum.

\paragraph{QuSpin cross-validator.}
Twenty-point $h$-sweep of the 4-site TFIM yields 320 eigenvalue
comparisons; 4-site Heisenberg full spectrum; 5-point $\Delta$-sweep
of XXZ.

\paragraph{High-precision validator.}
Diagonalizes TFIM, Heisenberg, and Kitaev chain at 50 decimal digits
using \texttt{mpmath} and compares against double-precision LAPACK.
Disagreement at the level of $10^{-15}$ rules out any systematic
double-precision pathology.

\paragraph{Sparse-vs-dense validator.}
Compares the lowest five eigenvalues from (ARPACK Lanczos) 
\texttt{scipy.sparse.linalg.eigsh}
against (LAPACK direct) \texttt{numpy.linalg.eigh} on
five models at $N=6$. Lanczos uses a Krylov-subspace iteration that
shares no code with the dense path; agreement provides
algorithm-independent confirmation of the lowest spectrum.

\paragraph{Jordan-Wigner validator.}
For models that are equivalent to free fermions under JW, the entire
spectrum is given by single-particle dispersion. Tests:
\textbf{(i)} TFIM PBC ground state from Bogoliubov de Gennes
dispersion,
\textbf{(ii)} XX chain (XXZ at $\Delta=0$) full $2^N$ spectrum,
\textbf{(iii)} free fermion OBC full spectrum,
\textbf{(iv)} Kitaev chain BdG cross-check (informational).

\paragraph{Spectral sum-rule validator.}\mbox{}\\
For each model, the validator checks four basis-independent invariants
against the spectrum:
\begin{equation}
\mathrm{tr}(H^k) \stackrel{?}{=} \sum_n E_n^k,
\quad k=1,2,3;
\qquad
\|H\|_2 \stackrel{?}{=} \max_n |E_n|.
\end{equation}
These are exact identities; any deviation indicates a computational
error.

% \paragraph{Spectral sum-rule validator.}
% For each model, checks four basis-independent invariants against the
% spectrum:
% \begin{equation}
% \mathrm{tr}(H^k) \stackrel{?}{=} \sum_n E_n^k, \quad k=1,2,3;
% \qquad \|H\|_2 \stackrel{?}{=} \max_n |E_n|.
% \end{equation}
% These are exact identities; deviation indicates a computational error.
%
\paragraph{Orthonormality validator.}
Checks that the eigenvector matrix~$V$ satisfies
$V^\dagger V = \mathbf{I}$, $V V^\dagger = \mathbf{I}$, and
$V D V^\dagger = H$ across nine models.

\paragraph{Unitarity validator.}
Builds $U(t) = V e^{-iDt} V^\dagger$ from the spectral decomposition
and checks (a) unitarity $UU^\dagger = \mathbf{I}$, (b) the group law
$U(t_1) U(t_2) = U(t_1 + t_2)$, and (c) agreement with
\texttt{scipy.linalg.expm} at $t=1$.

\paragraph{Conservation-law validator.}
For pairs $(H, S)$ where $[H,S]=0$ is expected, verifies the commutator
norm is $<10^{-12}$ \emph{and} that $S$ is block-diagonal in any
eigenbasis of~$H$ within the degenerate subspaces. Includes a
\emph{must not conserve} test (TFIM does not conserve total $S^z$) to
verify the validator distinguishes real from spurious symmetries.

\paragraph{Symmetry-sector validator.}
Projects $H$ onto each eigenspace of the conserved quantity, diagonalizes
each block independently, and verifies the union of sector spectra
equals the full spectrum. Tested for Heisenberg/$S^z$, TFIM/parity,
XXZ/$S^z$.

\paragraph{Thermal validator.}
Checks high-temperature ($\beta \to 0$) and low-temperature ($\beta \to \infty$)
limits for four spin models. At $\beta=0$, $\langle O \rangle \to
\mathrm{tr}(O)/d$ and $Z \to d$; at $\beta \to \infty$, thermal
averages converge to ground-state expectations.

\paragraph{Finite-size-scaling validator.}
Verifies that $E_0/N$ trends toward known thermodynamic limits as
$N$ grows: Heisenberg PBC $\to 1/4 - \ln 2 \approx -0.4431$ (Bethe
ansatz), free fermion half-filled $\to -2t/\pi \approx -0.6366$, TFIM
at criticality $\to -4/\pi \approx -1.2732$.

\paragraph{Error-injection validator.}
Injects six classes of error and verifies the framework detects each:
non-Hermitian term; matrix corruption (eigenvalues against the wrong
matrix); a deliberately corrupt oracle (consensus must fail);
eigenvector perturbation (residual must amplify); SHA-256 tampering
(load must fail); swapped eigenvectors of non-degenerate states
(residual must reveal mismatch). \emph{A verification framework that
fails its own injection tests cannot be trusted.}

% =============================================================================
\section{Results}
\label{sec:results}
% =============================================================================

We report results from a single end-to-end run on a Linux x86\_64
workstation (Python 3.14.4, NumPy 2.4.4, SciPy 1.17.1, mpmath 1.4.1,
QuSpin 1.0.1). The total wall-clock time was 5.92~s for the pytest
suite plus 5.16~s for the thirteen validators. The full result tree
(JSON, figures, hashed certificates) is included in the supplementary
archive.

\subsection{Headline numbers}

\begin{itemize}
\item Pytest unit + integration tests: \textbf{53 / 53 passed}.
\item Validator individual tests: \textbf{81 / 81 passed}.
\item Maximum disagreement against QuSpin across 320 TFIM
      \(h\)-sweep comparisons:
      \[
      \mathbf{1.6 \times 10^{-14}}
      \]
% \item Maximum disagreement against QuSpin (320 comparisons in TFIM
      % $h$-sweep): \textbf{$1.6 \times 10^{-14}$}.
\item Maximum disagreement against \texttt{mpmath} 50-digit
      reference: \textbf{$1.6 \times 10^{-15}$}.
\item Maximum spectral sum-rule error across 11 models:
      \textbf{$2.0 \times 10^{-13}$}.
\item Maximum orthonormality error across 9 models:
      \textbf{$3.6 \times 10^{-15}$}.
\item Maximum unitarity error across 7 models:
      \textbf{$2.2 \times 10^{-15}$}.
\item Error injection: \textbf{6 / 6 detected}.
\end{itemize}

\begin{figure}[t]
\centering
\includegraphics[width=0.85\textwidth]{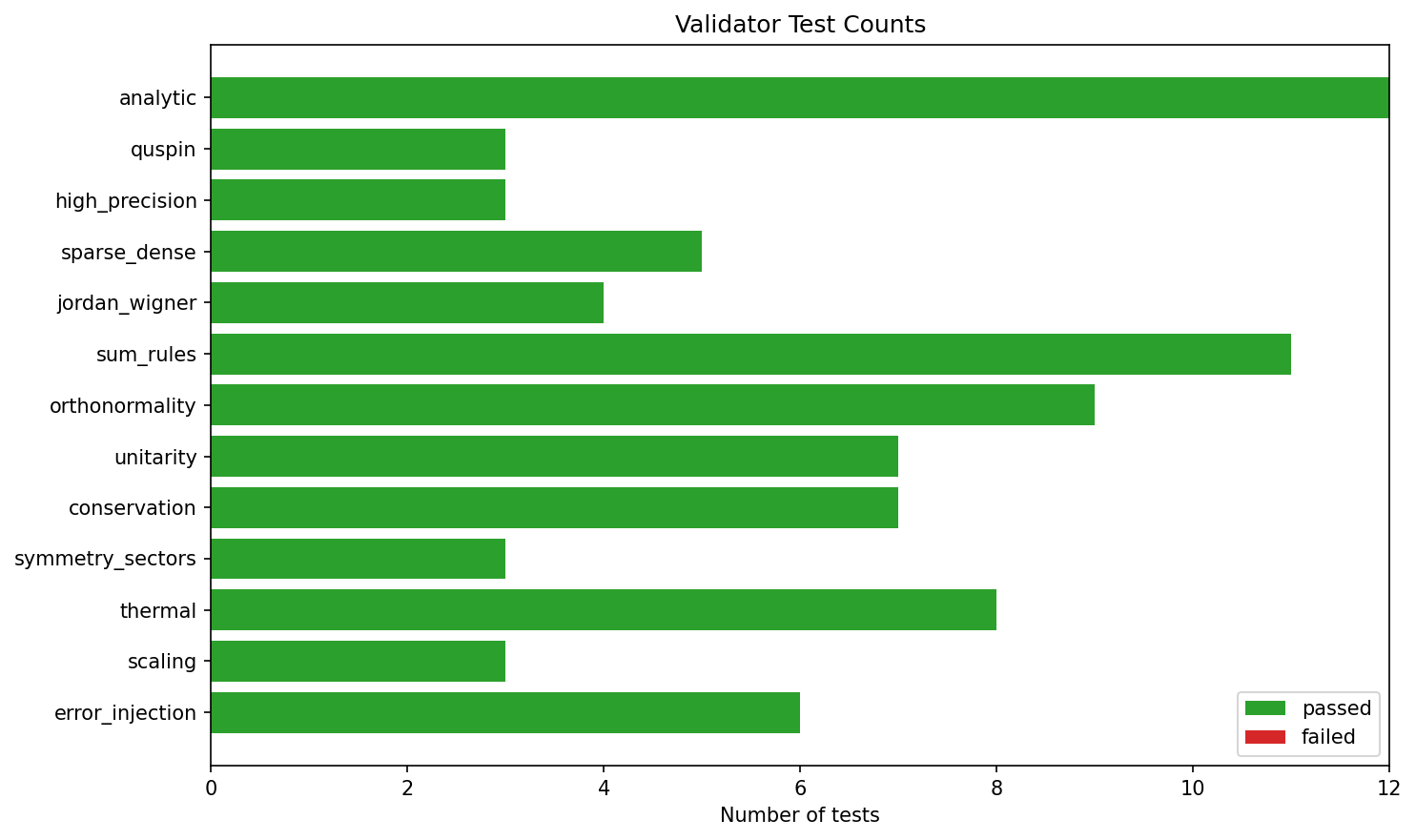}
\caption{Validator test counts. Each bar shows the number of individual
tests in one validator; all 81 tests pass on the run reported here.}
\label{fig:summary}
\end{figure}

These numbers establish that the framework's claims are not aspirational:
on every model in the registry, every validator, every comparison, the
errors are at or below the limits set by IEEE 754 double precision.
Figure~\ref{fig:summary} visualizes the full pass/fail count.

\subsection{Quantitative agreement with QuSpin}

The most important external cross-check is against
QuSpin~\cite{Weinberg2017,Weinberg2019}, an independently implemented
peer-reviewed ED package. Its implementation is entirely independent of
\textsc{certify-ed}.

We compare the full sixteen-dimensional spectrum of the 4-site
transverse-field Ising model at twenty values of the transverse field
\(h \in [0.1, 2.0]\), yielding 320 eigenvalue comparisons.

The maximum absolute disagreement across all comparisons is
\[
1.60 \times 10^{-14}.
\]

The mean disagreement is
\[
\bar{\Delta}_{\mathrm{TFIM}} = 1.6 \times 10^{-15},
\]
which also lies within the expected double-precision noise floor.

% The most important external cross-check is against
% QuSpin~\cite{Weinberg2017,Weinberg2019}, an established peer-reviewed
% ED package implemented entirely independently of \textsc{certify-ed}.
% We compare the full sixteen-dimensional spectrum of the 4-site
% transverse-field Ising model at twenty values of the transverse field
% $h \in [0.1, 2.0]$, yielding 320 eigenvalue comparisons. The maximum absolute disagreement across all 320 comparisons is
% $1.60 \times 10^{-14}$. The mean disagreement,
% $\bar{\Delta}_{\mathrm{TFIM}} = 1.6 \times 10^{-15}$,
% also lies within the expected double-precision noise floor.
%
% The maximum absolute disagreement across all 320 comparisons is
% $1.60 \times 10^{-14}$, with mean disagreement
% $\bar{\Delta}_{\mathrm{TFIM}} = 
% 1.6 \times 10^{-15}$, both within the
% expected double-precision noise floor. 
The 4-site Heisenberg full
spectrum agrees to $3.3 \times 10^{-16}$, and the XXZ anisotropy sweep
over $\Delta \in \{0, 0.5, 1.0, 1.5, 2.0\}$ agrees to $2.0 \times 10^{-15}$.

\subsection{Spectral sum rules}

\begin{figure}[t]
\centering
\includegraphics[width=0.95\textwidth]{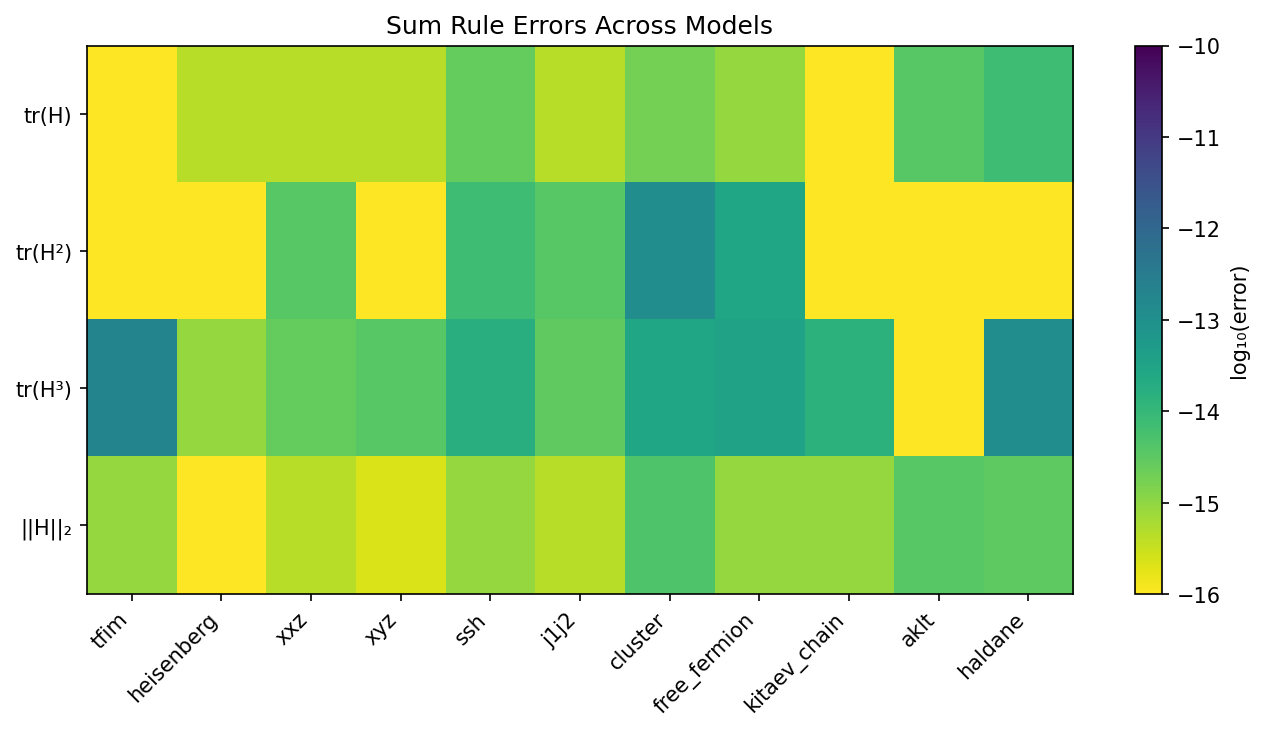}
\caption{Spectral sum-rule errors across eleven models. Color encodes
$\log_{10}$ of the absolute error between the trace identity and the
sum of eigenvalue powers. All errors lie between $10^{-16}$ and $10^{-13}$.}
\label{fig:sumrules}
\end{figure}

For a Hermitian matrix $H$ with eigenvalues $\{E_n\}$, the identities
$\mathrm{tr}(H^k) = \sum_n E_n^k$ are exact regardless of basis. We
test $k=1, 2, 3$ across eleven models, plus the operator-norm identity
$\|H\|_2 = \max_n |E_n|$. Figure~\ref{fig:sumrules} shows the absolute
error for each (model, sum rule) pair on a $\log_{10}$ scale; all
errors fall between $10^{-16}$ (machine precision) and $10^{-13}$
(roundoff scaling with $\|H\|^3$ for the cubed trace).

\subsection{Finite-size scaling against thermodynamic limits}

\begin{figure}[t]
\centering
\includegraphics[width=\textwidth]{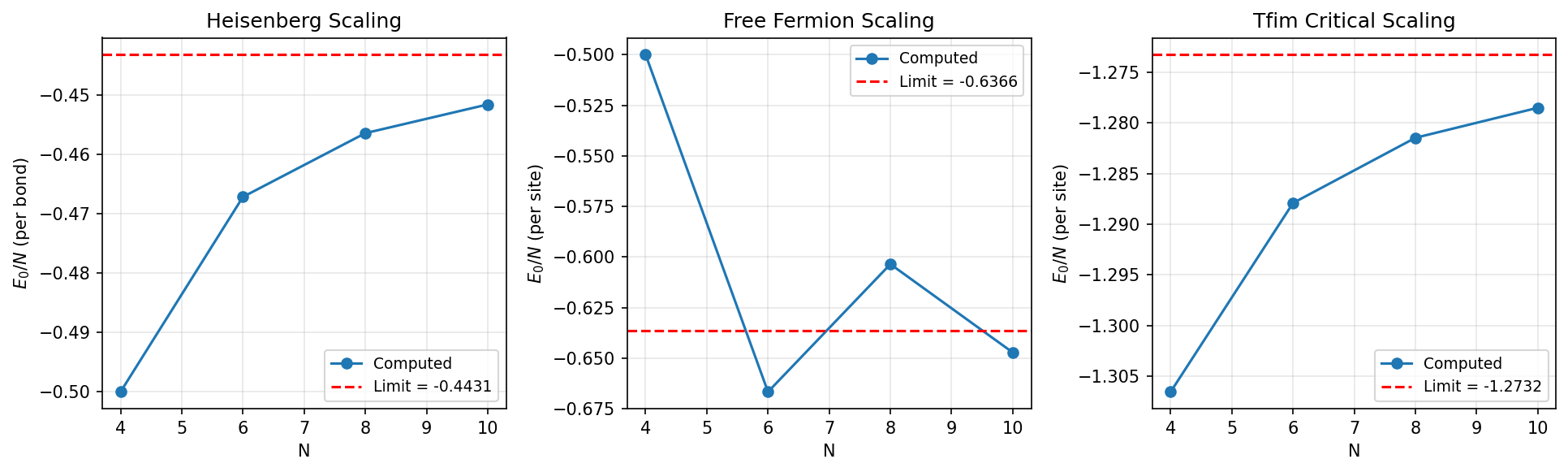}
\caption{Finite-size scaling toward thermodynamic limits. Left:
Heisenberg PBC ground-state energy per bond approaching the Bethe
ansatz value $1/4 - \ln 2$. Center: free fermion half-filled $E_0/N
\to -2t/\pi$ (oscillations are shell-filling effects). Right:
TFIM at criticality $E_0/N \to -4/\pi$.}
\label{fig:scaling}
\end{figure}

We check that $E_0/N$ on three integrable models converges toward
its thermodynamic limit as the system size~$N$ increases. For
Heisenberg PBC, the Bethe-ansatz ground state energy per bond is
$1/4 - \ln 2 \approx -0.4431$ in the convention where
$H = J\sum \mathbf{S}_i \cdot \mathbf{S}_j$. Free fermions at
half filling have $E_0/N \to -2t/\pi \approx -0.6366$ from the
band integral. The TFIM at the critical point $h/J = 1$ has
$E_0/N \to -4/\pi \approx -1.2732$~\cite{Pfeuty1970}. Figure~\ref{fig:scaling}
shows the convergence: by $N=10$ the deviations are below $1\%$ for
Heisenberg and TFIM critical, and below $2\%$ for free fermion (the
oscillations in the latter are well-known shell-filling effects).

\subsection{Error-injection self-test}

\begin{figure}[t]
\centering
\includegraphics[width=0.85\textwidth]{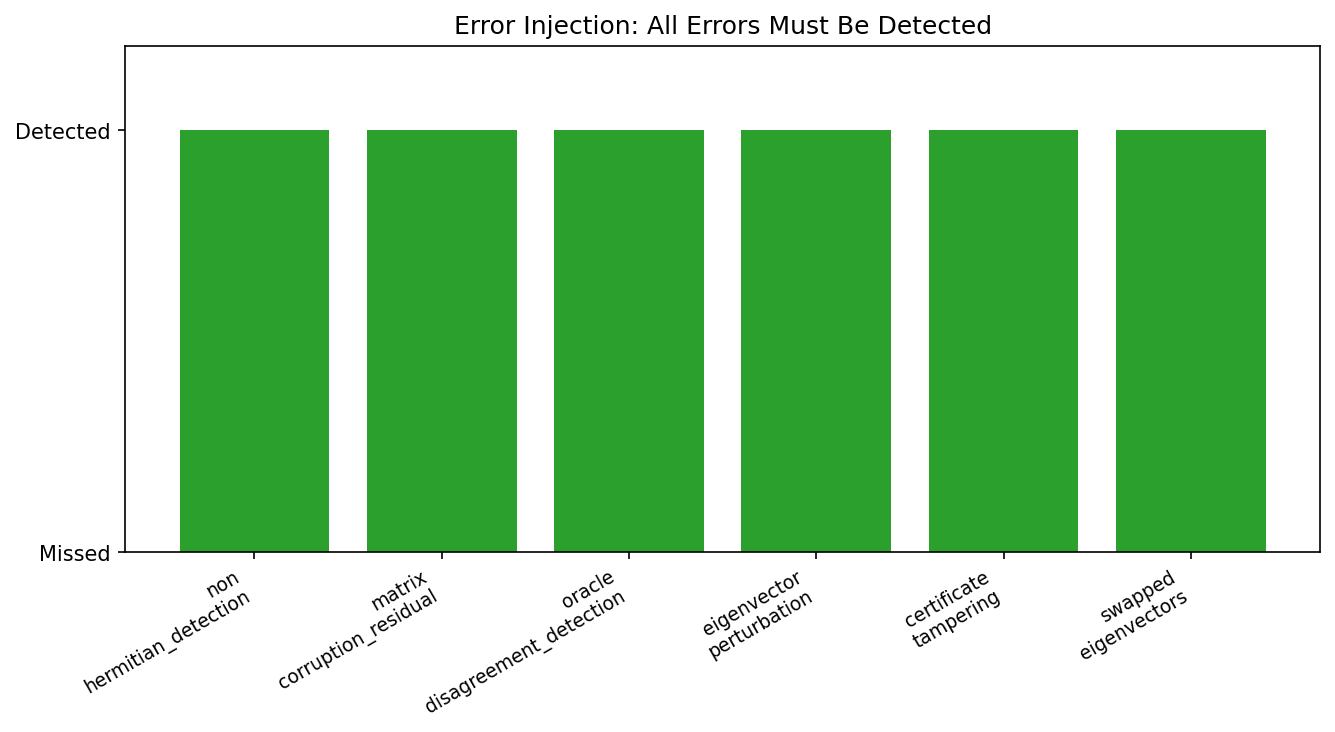}
\caption{Error-injection self-test. Six classes of error are injected
into the pipeline; each must be detected by at least one validator
or internal check. All six are detected.}
\label{fig:errinj}
\end{figure}

A verification framework that cannot detect injected errors gives
false confidence. We inject six classes of error and verify the
framework catches each (Figure~\ref{fig:errinj}):

\begin{enumerate}
% \item Non-Hermitian perturbation: $H \to H + i\epsilon\,(|0\rangle\langle 1|)$.
\item Non-Hermitian perturbation:
\[
H \to H + i\epsilon |0\rangle\!\langle 1|.
\]
      The Hermiticity check raises immediately at $\epsilon = 0.1$.
\item Matrix corruption: diagonalize $H + \delta H$ with random
      Hermitian noise of norm \[
      \|\delta H\|_F \sim 10^{-5}
      \], then
      compute residuals against the original $H$. Maximum residual
      jumps from $10^{-15}$ to $\sim 10^{-5}$, correctly flagging
      the inconsistency.
\item Oracle disagreement: a deliberately mis-shifted oracle is added
      to the multi-oracle ensemble. Consensus check fails at the
      $10^{-3}$ level, exceeding the $10^{-10}$ tolerance.
\item Eigenvector perturbation: state $|0\rangle$ is mixed with state
      $|1\rangle$ at amplitude $\alpha = 0.01$. Residual amplifies by
      a factor of $\sim 10^{14}$ relative to the unperturbed value.
\item Certificate tampering: a single eigenvalue is overwritten in the
      saved JSON. SHA-256 mismatch raises \texttt{ValueError} on load.
\item Swapped eigenvectors: $|0\rangle$ and $|1\rangle$ are interchanged
      while keeping the eigenvalues in original order. Residuals against
      the matched eigenvalues grow to $\sim |E_1 - E_0|$, revealing the
      mismatch.
\end{enumerate}

All six injections are detected. This is the strongest evidence that
the framework's internal checks have practical sensitivity.

% =============================================================================
\section{Discussion and Limitations}
\label{sec:discussion}
% =============================================================================

\paragraph{Multi-oracle is a soft check, not a guarantee.}
The three default oracles all link against LAPACK. A deep LAPACK bug
shared across DSYEVD and DSYEVR would not be caught by oracle consensus
alone. What multi-oracle does catch in practice is wrapper-level
inconsistencies between NumPy and SciPy (which have occurred
historically), algorithm-specific instabilities (DSYEVR's relatively
robust representations and DSYEVD's divide-and-conquer respond
differently to ill-conditioned matrices), and silent fallbacks in
either wrapper. The high-precision validator complements multi-oracle
by using a separate arithmetic stack (\texttt{mpmath}); the JW and
analytic validators complement both by using closed-form expressions
that bypass LAPACK entirely. The defense in depth is the point.

\paragraph{System sizes are limited by the dense path.}
We diagonalize matrices up to $\sim 2^{14}$. For larger systems, users
need symmetry-resolved sparse Lanczos as in QuSpin or XDiag.
\textsc{certify-ed} is intended to be applied either (a) to the small
benchmark systems where ED is the reference method, or (b) to the
small symmetry sectors of larger systems decomposed by
QuSpin/XDiag.

\paragraph{Rigorous interval enclosures are out of scope here.}
The framework reports residuals and consensus disagreements these
are evidence of correctness, not mathematical proof of correctness in
the verified-numerics sense. Follow-up work, currently in preparation,
develops verified eigenvalue enclosures for the symmetry-block-decomposed
Hamiltonians that arise naturally in many-body lattice models, with
meticulous control on the propagation of finite-precision errors from
the block decomposition into per-block enclosures. That direction
uses a different mathematical machinery (interval arithmetic,
fixed-point verification, perturbation bounds for invariant subspaces)
and is naturally a separate paper. For the workflows that motivate
the present work producing trustworthy ED output for downstream
theoretical claims the empirical validation framework presented
here is already a substantial improvement over current practice.

\paragraph{What this framework does not address.}
The framework verifies that, given a Hamiltonian~$H$, the spectrum and
eigenvectors returned are correct numerical solutions of $H\psi = E\psi$.
It does \emph{not} verify that $H$ is the correct Hamiltonian for a
given physical problem that is the user's responsibility, and a
verified ED of the wrong Hamiltonian is just as useless as an
unverified ED of the right one. The model registry is intended as
a check on the user's Hamiltonian assembly: a user who builds a TFIM
through the symbolic interface can compare against the canned
\texttt{build\_tfim} as one independent confirmation of model identity.

% =============================================================================
\section{Conclusion}
\label{sec:conclusion}
% =============================================================================

We have presented \textsc{certify-ed}, a verification framework for
exact diagonalization of quantum many-body Hamiltonians. The framework
imports the verified-numerics tradition of multi-path validation,
arbitrary-precision reference computation, and tamper-evident output
into the workflow of computational many-body physics, where these
practices are not yet standard. Across sixteen physical models and
thirteen independent validation layers, the reference implementation
passes all 81 individual validation tests, agrees with QuSpin to
$1.6 \times 10^{-14}$ over 320 eigenvalue comparisons, agrees with
50-digit \texttt{mpmath} reference values to $1.6 \times 10^{-15}$,
and detects all six classes of injected error. The total run time of
the full pipeline is under thirty seconds.

The framework is complementary to existing ED packages, not a
replacement. We hope it stimulates a discussion in the many-body
community about what evidence accompanies an ED result when that
result underpins a published claim. The current de facto standard
implicit trust in LAPACK plus informal cross-checking is, we
suggest, ripe for upgrade.

\paragraph{Code and data availability.}
The full source, test suite, validators, and master pipeline runner
are released under MIT license\cite{Sarang2026}. The hashed certificates for the
results reported here are included in the supplementary archive.

\paragraph{Acknowledgments.}
The author thanks the QuSpin and XDiag developers for their
peer-reviewed open-source codes, which served as ground-truth references
during development. 

% =============================================================================
% References
% =============================================================================

% \begin{verbatim}
% \bibliography{your_bibtex_file}
% \end{verbatim}
% at the end of your document. If you are not using our LaTeX template, please still use our bibstyle as
% \begin{verbatim}
% \bibliographystyle{SciPost_bibstyle}
% \end{verbatim}
% in order to simplify the production of your paper.
% \end{appendix}
%


% Provide your bibliography here. You have two options:

%%% FIRST OPTION
% Write your entries here directly, following the example below, including:
% Author(s), Title, Journal Ref. with year in parentheses at the end, followed by the DOI number.

%\begin{thebibliography}{99}
%\bibitem{1931_Bethe_ZP_71} H. A. Bethe, {\it Zur Theorie der Metalle. i. Eigenwerte und Eigenfunktionen der linearen Atomkette}, Zeit. f{\"u}r Phys. {\bf 71}, 205 (1931), \doi{10.1007\%2FBF01341708}.
%\bibitem{arXiv:1108.2700} P. Ginsparg, {\it It was twenty years ago today... }, \url{http://arxiv.org/abs/1108.2700}.
%\end{thebibliography}

% Use your bibtex library, formatted by the SciPost style file.
% \bibliography{SciPost_Example_BiBTeX_File.bib}

\begin{thebibliography}{99}

\bibitem{Weinberg2017}
P.~Weinberg and M.~Bukov,
\emph{QuSpin: a Python Package for Dynamics and Exact Diagonalisation
of Quantum Many Body Systems part I: spin chains},
SciPost Phys.\ \textbf{2}, 003 (2017).

\bibitem{Weinberg2019}
P.~Weinberg and M.~Bukov,
\emph{QuSpin: a Python Package for Dynamics and Exact Diagonalisation
of Quantum Many Body Systems. Part II: bosons, fermions and higher spins},
SciPost Phys.\ \textbf{7}, 020 (2019).

\bibitem{Wietek2025}
A.~Wietek \emph{et al.},
\emph{XDiag: Exact Diagonalization for Quantum Many-Body Systems},
arXiv:2505.02901 (2025), submitted to SciPost Physics Codebases.

\bibitem{Bauer2011}
B.~Bauer \emph{et al.},
\emph{The ALPS project release 2.0: open source software for strongly
correlated systems},
J.\ Stat.\ Mech.\ (2011) P05001.

\bibitem{Rump2001}
S.~M.~Rump,
\emph{Computational error bounds for multiple or nearly multiple eigenvalues},
Linear Algebra Appl.\ \textbf{324}, 209 (2001).

\bibitem{Rump1999}
S.~M.~Rump,
\emph{INTLAB: INTerval LABoratory},
in T.~Csendes (ed.), Developments in Reliable Computing,
Kluwer Academic Publishers, 77 (1999).

\bibitem{Miyajima2010}
S.~Miyajima,
\emph{Fast enclosure for all eigenvalues in generalized eigenvalue problems},
J.\ Comput.\ Appl.\ Math.\ \textbf{233}, 2994 (2010).

\bibitem{vanderHoeven2017}
J.~van der Hoeven,
\emph{Efficient Certification of Numeric Solutions to Eigenproblems},
in MACIS 2017, LNCS \textbf{10693}, 81 (2017).

\bibitem{Hashimoto2022}
K.~Hashimoto, K.~Morikuni, T.~Sogabe and S.-L.~Zhang,
\emph{Verified eigenvalue and eigenvector computations using complex
moments and the Rayleigh-Ritz procedure for generalized Hermitian
eigenvalue problems},
arXiv:2110.01822 (2022).

\bibitem{Huber2020}
S.~P.~Huber \emph{et al.},
\emph{AiiDA 1.0, a scalable computational infrastructure for automated
reproducible workflows and data provenance},
Sci.\ Data \textbf{7}, 300 (2020).

\bibitem{Pfeuty1970}
P.~Pfeuty,
\emph{The one-dimensional Ising model with a transverse field},
Ann.\ Phys.\ (N.Y.) \textbf{57}, 79 (1970).

\bibitem{Bethe1931}
H.~Bethe,
\emph{Zur Theorie der Metalle. I.~Eigenwerte und Eigenfunktionen der
linearen Atomkette},
Z.\ Phys.\ \textbf{71}, 205 (1931).

\bibitem{Lieb1961}
E.~Lieb, T.~Schultz and D.~Mattis,
\emph{Two soluble models of an antiferromagnetic chain},
Ann.\ Phys.\ (N.Y.) \textbf{16}, 407 (1961).

\bibitem{Affleck1987}
I.~Affleck, T.~Kennedy, E.~H.~Lieb and H.~Tasaki,
\emph{Rigorous results on valence-bond ground states in antiferromagnets},
Phys.\ Rev.\ Lett.\ \textbf{59}, 799 (1987).

\bibitem{Kitaev2001}
A.~Y.~Kitaev,
\emph{Unpaired Majorana fermions in quantum wires},
Phys.\ Usp.\ \textbf{44}, 131 (2001).

\bibitem{Anderson1987}
P.~W.~Anderson,
\emph{The resonating valence bond state in $\mathrm{La_2 CuO_4}$ and
superconductivity},
Science \textbf{235}, 1196 (1987).

\bibitem{Suzuki1971}
M.~Suzuki,
\emph{Relationship between d-Dimensional Quantal Spin Systems and
(d+1)-Dimensional Ising Systems: Equivalence, Critical Exponents and
Systematic Approximants of the Partition Function and Spin Correlations},
Prog.\ Theor.\ Phys.\ \textbf{56}, 1454 (1976).

\bibitem{Sarang2026}
S.~Vehale and R.~Goel,
\emph{sarang-kernel/CERTIFY-ED: v1.0.0},
Zenodo (2026),
DOI: 10.5281/ZENODO.20066565,

\end{thebibliography}

%%%%%%%%%% END TODO: BIBLIOGRAPHY

\end{document}